\documentclass{iaus}
\usepackage{graphicx}

\newcommand{\msun}{$M_\odot$}

\newcommand{\hii}{H\,{\sc ii}\rm}
\newcommand{\hei}{He\,{\sc i}\rm}

\newcommand{\siii}{[S\,{\sc iii}]}
\newcommand{\siv}{[S\,{\sc iv}]}
\newcommand{\nii}{[N\,{\sc ii}]}

\newcommand{\oiii}{[O\,{\sc iii}]}
\newcommand{\oiicoll}{[O\,{\sc ii}]}

\newcommand{\sii}{[S\,{\sc ii}]}
\newcommand{\ariii}{[Ar\,{\sc iii}]}
\newcommand{\neiii}{[Ne\,{\sc iii}]}

\newcommand{\cii}{C\,{\sc ii}\rm}
\newcommand{\oii}{O\,{\sc ii}\rm}

\newcommand{\neii}{[Ne\,{\sc ii}]}

\newcommand{\te}{$T_e$}

\newcommand{\lin}{$\,\lambda$}
\newcommand{\llin}{$\,\lambda\lambda$}

\title[Massive stars in the nuclei and arms of spirals] %% give here short title %%
{Massive stars in the nuclei and arms of spirals}

\author[Fabio Bresolin]   %% give here short author list %%
{Fabio Bresolin}

\affiliation{Institute for Astronomy, University of Hawai'i\\ 2680 Woodlawn Drive, 96822 Honolulu, Hawai'i, USA\\ email: {\tt bresolin@ifa.hawaii.edu} \\}

\pubyear{2008}
\volume{250}  %% insert here IAU Symposium No.
\pagerange{119--126}
\jname{Massive Stars as Cosmic Engines}
\editors{F. Bresolin, P.A. Crowther \& J. Puls, eds.}
\begin{document}

\maketitle

\begin{abstract}
Many of the properties of massive stars in external galaxies, such as chemical compositions, mass functions, and ionizing fluxes, can be derived from the 
study of the associated clouds of ionized gas. Moreover, the signatures of Wolf-Rayet stars are often detected in the spectra of extragalactic \hii\/ regions. This paper reviews some aspects of the recent work on the massive star content of nearby spiral galaxies, as inferred from the analysis of giant \hii\/ regions. Particular attention is given to regions of high metallicity, including nuclear hot spots, and to the chemical abundance comparison between supergiant stars and ionized gas.

\end{abstract}

\firstsection % if your document starts with a section,
\section{Introduction}

Hot and massive stars are among the most conspicuous objects visible in the arms and the nuclei of spiral galaxies. 
However, there is more than meets the eye. Young massive stars in external galaxies can be detected well beyond the optical galactic boundary defined by the isophotal radius.
In the nearby NGC~2403, \cite[Davidge~(2007)]{} has found young main sequence stars out to 8 disk scale lengths (16 kpc) from the galactic center. The GALEX mission has revealed the presence of UV-bright complexes, i.e.~clusters of young massive stars, in the outskirts of about 1/4 of the nearby spiral galaxy sample it has observed. For example, in M83 \cite[Thilker et al.~(2005)]{} have detected UV-bright knots out to 4 disk scale lengths. The star formation rate in these external regions is fairly low, as also indicated by the fact that only 5-10\% of the UV knots are detected in H$\alpha$ (\cite[Zaritsky \& Christlein~2007]{}). The oxygen abundance in the outer \hii\/ regions in M83 has been determined by \cite[Gil de Paz~(2007)]{} to be around 10\% of the solar value, while the center of the galaxy reaches approximately twice the solar value (\cite[Bresolin et al. 2005; Pellerin \& Robert~2007]{}). Finally, what appear to be intergalactic \hii\/ regions have been found out to 30 kpc from their nearest galaxy (\cite[Ryan-Weber et al.~2004]{}).

A large amount of additional work needs to be done on the properties of massive stars in the outer regions of spiral galaxies, in order to better understand the mechanisms that lead to their formation, and in general to learn more about the evolution of galaxies, for example about the relationship between neutral gas surface density and star formation rate, or the processes that lead to the build-up of the chemical elements. This review focuses on more `normal' sites of massive star formation, represented by giant \hii\/ regions located in the optically bright portions of spiral disks. Since the line emission from these 
objects is directly linked to the ionization provided by their massive star content, the study of giant \hii\/ regions in nearby galaxies, as well as in distant star-forming galaxies,
provides a wealth of information on the properties of the embedded clusters of ionizing stars (mass functions, hierarchical structuring), of the individual massive stars (initial mass functions, ionizing properties) and of the ionized gas (chemical abundances). For brevity's sake, only classical nebulae, photoionized by clusters of hot O and Wolf-Rayet stars, will be considered here, as space does not allow to cover other sources of line emission found in spiral galaxies, such as Seyfert nuclear activity, LINERS or supernova remnants.
Also of particular interest, and the concluding topic of this article, are the rings of circumnuclear star formation that are present in a relatively large fraction of spiral galaxies in the nearby universe.

Some of the most interesting developments in our understanding of massive star formation and galaxy evolution derive from a multi-wavelength observational approach in the study of galaxy properties. In particular, recent space missions, such as GALEX in the ultraviolet and Spitzer in the mid- and far-infrared, have provided high-sensitivity access to star formation rate estimators, such as the UV stellar continuum and the monochromatic IR emission, that complement more traditional optical means, such as the optical recombination lines (H$\alpha$ in particular). The cross-correlation between these different star formation rate estimators, obtained locally in galaxies within a few tens of Mpc (\cite[Schmitt et al.~2006]{}), is crucial, for example, to gain confidence about the results obtained for high-redshift star-forming galaxies. Recent results from the Spitzer Infrared Nearby Galaxies Survey (SINGS)  are particularly interesting in this context. \cite[Calzetti et al.~(2007)]{} and \cite[Kennicutt et al.~(2007)]{} have investigated the use of the 24\,$\mu$m emission as a reliable star formation rate indicator in galaxies. Looking at both the optical and infrared line emission, \cite[Prescott et al.~(2007)]{} concluded that the fraction of highly obscured large star-forming regions in spiral disks is small, less than 4\%\/, implying that optical studies that rely on the H$\alpha$ emission to trace massive stars do not miss the bulk 
of massive star formation. This conclusion `justifies' the remaining part of this review, in which we look at massive star properties in connection with optically bright \hii\/ regions. It is also worth mentioning the existence of an additional project, the Survey for Ionization in Neutral Gas Galaxies (SINGG), in which the massive star-forming regions, detected from their H$\alpha$ emission, are studied in a radio-selected galaxy sample, providing a bias-free view of star-forming galaxies in the nearby universe (\cite[Meurer et al.~2006; Hanish et al.~2006]{}).\\

In the following sections I will briefly discuss some recent findings concerning: (1)~luminosity and mass functions of \hii\/ regions and star clusters; (2)~Wolf-Rayet star content, chemical abundance studies and ionizing fluxes in metal-rich \hii\/ regions; (3)~chemical abundance comparisons between blue supergiants and ionized gas; (4)~circumnuclear regions of massive star formation in spiral galaxies.

\section{The power of luminosity functions}
 
One of the simplest techniques used for the analysis of extragalactic \hii\/ regions and star clusters, the construction of luminosity functions, can provide powerful conclusions on the properties of the evolving populations of massive stars in other galaxies.

\hii\/ region luminosity functions have been studied in detail for about two decades. Recently, \cite[Bradley et al. (2006)]{} have published a composite luminosity function, modeled, as is customary, with a power law of the form $N(L)dL \propto L^{-\alpha}dL$, comprising approximately 18,000 \hii\/ regions in 56 spiral galaxies, and offering a clear view of the fact that the slope, $\alpha\simeq2$, becomes shallower at lower H$\alpha$ luminosities, as already observed by \cite[Kennicutt, Edgar \& Hodge~(1989)]{}. This can be due to a transition from ionization-bounded to density-bounded nebulae at a well-defined H$\alpha$ luminosity (\cite[Beckman et al.~2000]{}), or to the presence of stochastic variations in the number of ionizing stars at low \hii\/ region luminosity (\cite[Oey \& Clarke~1998]{}). Alternatively, it can be interpreted as the convolution of the time-dependent H$\alpha$ luminosity due to an evolving massive star population with a lognormal cluster mass function (\cite[Dopita et al.~2006a]{}).

The luminosity functions of young clusters in spiral galaxies can also be modeled using a power law, with a slope similar to that found for the \hii\/ region luminosity function (\cite[Larsen~2002]{}). A break in the slope observed in M51 by \cite[Gieles et al.~(2006)]{} has been interpreted as due to a truncation of the cluster mass function above $10^5$~\msun.

It has been pointed out by \cite[Elmegreen~(2006)]{} that the observed power-law form of the young cluster mass function in spiral galaxies, $N(M)dM \propto M^{-2}dM$, derives from the fact that the composite, system-wide stellar Initial Mass Function (IMF) of galaxies appears to have a slope that is close to the Salpeter value, as is found in the constituent individual stellar clusters and OB associations. This is the expectation from models of scale-free, hierarchical distributions of young star structures and star-forming sites within galaxies, from kpc-scale down to the size of individual stellar clusters. Recent observational confirmations of this hierarchical structuring have been found by \cite[Bastian et al.~(2007)]{} in the galaxy M33, and by \cite[Elmegreen et al.~(2006)]{} in NGC~628.

\section{The utility of nebular spectra}
The optical and near-IR spectra of \hii\/ regions are dominated by emission lines due to the recombination of H and He atoms, and to forbidden transitions, due to collisional excitation, of several metal ions, some of the strongest lines being \oiicoll\lin3727, \oiii\llin4959,5007, \nii\llin6548,6583, \sii\llin6717,6731, \siii\llin9069,9532, \ariii\lin7135 and \neiii\lin3868. The study of line ratios allows to measure chemical compositions and the excitation state of the gas. The latter depends on the energy input from the massive stars that ionize the gas. The chemical abundances of the gas are expected to be virtually the same as for the young stars (modulo some effects that include depletion on dust grains and mixing at the stellar surfaces). Most of our current knowledge on the chemical composition of spiral galaxies, and of star-forming galaxies in general, including those at high-redshift, comes from emission-line studies of the ionized gas. Abundances from stellar spectra in external galaxies are more difficult to obtain. Some stellar results are discussed in connection with nebular abundances in section~5.
\\

The spectroscopic study of extragalactic \hii\/ regions is being used to constrain massive star parameters in different ways. Some examples are only briefly mentioned here:\\

\begin{itemize}
\item{the \hii\/ region chemical abundances provide key observational constraints for galactic evolution models, and therefore on the input parameters that characterize these models, such as the stellar IMF, the star formation rate and efficiency, the stellar yields, and the importance of gas inflows and outflows. As a recent example, \cite[Magrini et al.~(2007a)]{} have used the chemical abundances of \hii\/ regions, blue supergiants, planetary nebulae and RGB stars measured across the disk of M33,  to model the temporal evolution of the chemical abundance gradient in this galaxy.}\\

\item{as shown by \cite[Morisset et al.~(2004)]{}, the ionizing flux output predicted for massive stars by different stellar atmosphere codes ({\sc cmfgen}, {\sc wm}-basic, CoStar, {\sc tlusty}, {\sc fastwind}) can differ by orders of magnitude at high photon energy (above 40 eV). Modeling nebulae with the prescriptions for the spectral energy distribution obtained from the different codes can therefore provide useful indications on the actual ionizing continuum flux distribution of O stars (\cite[Simon-Diaz \& Stasinska~2008]{}).}\\

\item{stellar masses and effective temperatures of the ionizing stars can be estimated from optical and IR diagnostic diagrams, as recently shown by \cite[Dopita et al.~(2006b)]{} for Galactic ultracompact \hii\/ regions, using Spitzer-accessible line ratios, such as \neiii15.5$\mu$m/ \neii12.8$\mu$m and \siv10.5$\mu$m/\ariii9.0$\mu$m.}\\

\item{the upper limit of the stellar initial mass function can be estimated using measurements of the equivalent width of the H$\beta$ emission line in extragalactic \hii\/ regions that contain Wolf-Rayet star signatures (\cite[Pindao et al.~2002]{}). Together with reliable chemical abundance measurements, this method indicates that the cut-off mass of the stellar IMF is not peculiar at high metallicity (\cite[Bresolin~2005]{})}, against several suggestions of the contrary present in the literature. \\

\end{itemize}

\section{Metal-rich H\,{\footnotesize II} regions}

\medskip

\subsection{Wolf-Rayet stars}
Emission features from  Wolf-Rayet (W-R) stars, e.g.~the WN bump at 4660\,\AA, are frequently observed in high-metallicity \hii\/ regions (\cite[Bresolin, Kennicutt \& Garnett~2004]{}). This is well understood in terms of stellar evolution models, 
that show that the duration of the W-R phase and the W-R/O number ratio
in an evolving ensemble of massive stars increase with metallicity (see the discussion in the recent paper on the Geneva models with rotation and Starburst99 by \cite[Vazquez et al.~2007]{}). The growing number of deep spectra of extragalactic \hii\/ regions obtained in recent years shows that the CIII\,5696 \AA\/ line of late WC (WCL) stars is only found in the central regions of spiral galaxies, which, in virtue of the radial abundance gradients, are the most metal-rich. For example, WCL features are detected in \hii\/ regions in M101 only where the oxygen abundance is close to the solar value (\cite[Bresolin~2007]{}). This agrees with early results from \cite[Phillips \& Conti~(1992)]{} that WCL stars are preferentially found in metal-rich environments. \cite[Crowther et al.~(2002)]{} showed that the CIII\,5696 \AA\/ 
line strength increases with mass-loss rate, and therefore with metallicity.

A clear example of the effect of metallicity on the W-R star content is given by \cite[Hadfield et al.~(2005)]{}, who detected W-R features in a large fraction of \hii\/ regions in M83, a nearby metal-rich galaxy (\cite[Bresolin \& Kennicutt~2002]{}), with WC8-9 stars being the dominant types.
The WC/WN number ratio has been shown by many authors to increase with gas-phase metallicity (see \cite[Crowther~2007]{} for a recent review), in rough agreement with predictions by \cite[Eldridge \& Vink~(2006)]{}. It is worth mentioning here that in the study of this and similar trends with metallicity, most of the information on metallicities is derived from nebular studies. When comparing quantitative results with model predictions it is therefore important to remember that the reliability of the nebular abundances is still somewhat in question, especially at high metallicity (solar and above).

\subsection{Chemical abundances}
There are many additional situations in which an accurate knowledge of nebular abundances is crucial for a comparison with the model predictions, for example 
in the study of the metallicity dependence of the number ratio of type Ibc to type II supernovae (\cite[Prieto et al.~2007]{}), and the determination of the mass-metallicity relation for high-redshift galaxies (\cite[Erb et al.~2006]{}).
Somewhat surprisingly, however, despite the fact that the study of nebular abundances is quite a mature field, the abundance scale is still somewhat uncertain, by at least a factor of two at the high-metallicity end (approximately solar metallicity and above). Skipping the details (see reviews by \cite[Bresolin~2006]{} and \cite[Stasinska~2007]{}), we can focus here on two major observational projects related to the determination of nebular abundances that have been developed since the previous massive star symposium in Lanzarote: the determination of `direct' abundances in high-metallicity nebulae, and the measurement of metal recombination lines in Galactic and extragalactic \hii\/ regions.\\

\noindent
1. The classical method of chemical abundance analysis in \hii\/ regions is based on the detection of faint auroral lines, such as \oiii\lin4363 and \nii\lin5755, that correspond to collisionally excited transitions that take place from the second lowest excited level of metal ions to the lowest excited level. In combination with nebular lines of the same ions originating from transitions from the lowest excited level to the ground level (e.g.~\oiii\lin5007 and \nii\lin6583) one can obtain a direct measurement of the electron temperature \te, and therefore fix the value of the line emissivity, which is highly sensitive to \te.

In recent years this technique has been extended towards the central, metal-rich zones of spiral galaxies, where the increased cooling from line emission leads to a strong decrease in the auroral-to-nebular line ratios
 (\cite[Bresolin et al.~2005]{}; \cite[Bresolin~2007]{}). There are some theoretical expectations that the method could generate erroneous results (\cite[Stasinska~2005]{}), although observationally there is no indication yet that abundance biases are significant for \hii\/ regions with metallicity up to the solar value. If the auroral line method is reliable at high metallicity, then the oxygen abundances measured in the central regions of the most metal-rich nearby spiral galaxies lie in the range 12+log(O/H)=8.60\,--\,8.75 (\cite[Pilyugin, Thuan \& Vilchez~2006]{}), i.e.~close the solar value [adopting 12+log(O/H)$_\odot$=8.66 after \cite[Asplund et al.~2004]{}], and 2-3 times smaller than found earlier, based on indirect methods that rely on semi-empirical calibrations of `strong-line' diagnostics rather than on the direct measurement of the nebular electron temperature. An important consequence of the lower abundances is that the effective yields used in closed box galactic evolution models are reduced, again by factors of 2-3 (\cite[Bresolin, Kennicutt \& Garnett~2004]{}; \cite[Pilyugin, Thuan \& Vilchez~2007]{}).\\

\noindent
2. Metal recombination lines, in particular \oii\lin4651 and \cii\lin4267, have been measured in a number of Galactic and extragalactic \hii\/ regions (\cite[Peimbert et al.~2007]{}). The abundances derived from metal recombination lines have the advantage of being only mildly dependent on the value of the electron temperature (while collisionally excited lines depend exponentially on it). However, these lines are very faint, and this has insofar limited their use in extragalactic work to only a handful of targets. \cite[Garcia-Rojas \& Esteban~(2007)]{} summarize the work done on nebular metal recombination lines, and show that oxygen abundances derived from recombination lines are, on average, a factor of two larger than those derived from collisionally excited lines (i.e.~using the auroral line method). The origin of this abundance discrepancy is still object of debate. While Peimbert and collaborators invoke the presence of temperature fluctuations, alternative explanations have been proposed. In particular, \cite[Stasinska et al.~(2007)]{} have recently proposed an explanation based on the existence of abundance inhomogeneities, in the form of metal-rich droplets, rained down upon the interstellar medium of spiral disks following the explosion of supernovae. However, \cite[Lopez-Sanchez et al.~(2007)]{} question this interpretation, claiming that it is not really supported by the observational data.\\

In summary, the values one derives for the oxygen abundances of \hii\/ regions 
still depend on the method used. Choosing the `direct' method (based on collisionally excited lines) produces a `low abundance scale' (e.g.~Pilyugin \& Thuan~2005), that differs by a factor of 2-3 relative to the `high abundance scale' obtained from the use of metal recombination lines or photoionization models (e.g.~\cite[Kewley \& Dopita~2002]{}). This obviously has an immediate impact on the calibration of the so-called `strong line methods', used throughout the literature to estimate nebular chemical abundances in star-forming galaxies from the strength of the most prominent optical forbidden emission lines, such as in 
the widely used parameter $R_{23}$~=~(\oiicoll~+~\oiii)/H$\beta$.

\subsection{Radiation field}
The softening of the ionizing radiation and the decrease of the effective temperature of the ionizing stars of extragalactic \hii\/ regions with increasing metallicity  have been investigated since the early extragalactic nebular studies in the 1970's. Optical diagnostics of effective temperature, such as \hei\lin5876/H$\beta$ and \oiii\lin5007/H$\beta$, show an apparent softening of the radiation field at high metallicity (\cite[Dors \& Copetti~2003]{}; \cite[Bresolin, Kennicutt \& Garnett~2004]{}). Infrared diagnostics (e.g.~\neiii15.5$\mu$m/\neii12.8$\mu$m) confirm this finding (\cite[Rigby \& Rieke~2004]{}).
This has often given support to the idea that the upper mass cut-off of the stellar mass function is lowered, or that the IMF slope is steepened (\cite[Zhang et al.~2007]{}), in metal-rich environments. The softening, however, can be explained without invoking changes in the mass function when hot star models in non-LTE and including line blanketing are used in population synthesis models to predict the evolution of the emission line ratios (\cite[Smith, Norris \& Crowther~2002]{}). The consideration of extended bursts of star formation instead of single-burst episodes also helps to discard anomalous IMFs in metal-rich starbursts (\cite[Fernandes et al.~2004]{}).

\cite[Martin-Hernandez et al.~(2002)]{} attributed the decrease in the degree of ionization of neon with metallicity, observed in Galactic and extragalactic \hii\/ regions and in starbursts with the ISO satellite,  to a metallicity effect on the spectral energy distribution of the ionizing stars. A similar conclusion has been drawn by \cite[Morisset et al.~(2004)]{} and \cite[Mokiem et al.~(2004)]{}.

\cite[Rubin et al.~(2007)]{} have published one of the first tests between theoretical expectations and mid-IR observations  obtained with the IRS spectrograph aboard the Spitzer space telescope.
Since high-metallicity galaxies provide significant testbeds for theoretical stellar atmospheres, these authors have observed a number of \hii\/ regions across the disk of the metal-rich galaxy M83. When using different stellar atmosphere codes to produce nebular models, the best match with the observed emission line ratios \siv10.5$\mu$m/\siii18.7$\mu$m and \neiii15.6$\mu$m/\siii18.7$\mu$m is obtained with {\sc wm}-basic supergiant models. Interestingly, these authors, together with \cite[Simpson et al.~(2004)]{}, point out that modern stellar atmosphere codes are still generally unable to reproduce the observed high values of the Ne$^{++}$/O$^{++}$ ratio at low O$^{++}$/S$^{++}$, known as the \neiii\/ problem, and that was apparently considered solved in the mid-1990's with the first generation of non-LTE stellar atmosphere codes (\cite[Sellmaier et al.~1996]{}).
Further tests, in particular involving spectroscopic observations of both the ionizing source and the line-emitting gas of single-star \hii\/ regions compared to model nebulae, will help shed some light on the reliability of currently available model atmospheres of hot stars (\cite[Simon-Diaz et al.~2007]{}).

\section{Nebular vs. stellar abundances}

For a long time measuring the chemical abundances of the young populations 
in spiral galaxies has been done mainly via the analysis of optical and IR emission lines of the ionized regions surrounding hot stars. The more direct approach of analyzing single stars in other spirals has been mostly limited to bright supergiants in nearby galaxies of the Local Group, namely M33 and M31. In the last few years high signal-to-noise spectra of single supergiant stars have been obtained with 8m-class telescopes at distances of up to 7 Mpc (\cite[Bresolin et al.~2001]{}). At larger distances one can rely on the integrated spectrum of the ionizing clusters, for example in the UV, and deduce the chemical composition from the comparison with population synthesis models. One recent example of this `integrated' approach is the IR spectroscopic study of the super-star cluster in NGC~6946 by \cite[Larsen et al.~(2006)]{}. Using synthetic stellar spectra to reproduce the H- and K-band spectral features, which are dominated by 
red supergiants ($\sim$15 Myr of age), these authors have derived an oxygen abundance 12+log(O/H)=8.66 (i.e.~solar), with an indication of an $\alpha$/Fe element ratio enhanced by about 0.2 dex relative to solar. From the known nebular abundance gradient, I have obtained for the surrounding \hii\/ regions 12+log(O/H)=8.95, using strong-line methods calibrated in the high abundance scale, and 12+log(O/H)=8.55 using the low abundance scale (P method of Pilyugin). In NGC~1569 \cite[Larsen et al.~(2008)]{} have obtained a 0.25 dex discrepancy in oxygen content between stars and ionized gas.

Since both hot, massive stars and ionized gas sample the present-day chemical 
composition of galactic disks, we expect a good match between the results obtained from the 
two kinds of objects. The comparison can be complicated by a number of factors, including dust depletion in the gas, mixing at the stellar surface, and in general by the details of the chemical analysis that is done with different techniques and sets of spectral lines. The comparison is commonly restricted to the oxygen abundances, because this element can be measured relatively easily both in stars (B supergiants) and \hii\/ regions (typically the strongest metal emission lines). 
Thus far, comparing single-star chemical abundances with those in ionized nebulae has been carried out in a very restricted number of spiral galaxies. The best example in the northern hemisphere is arguably M33, while in the southern sky the best case is offered by NGC~300, a galaxy that is very similar to M33 in appearance, but located at approximately 2.5 times the distance ($\sim$2 Mpc).

\subsection{NGC~300}
\begin{figure}[b]
% \vspace*{-2.0 cm}
\begin{center}
 \includegraphics[width=\textwidth]{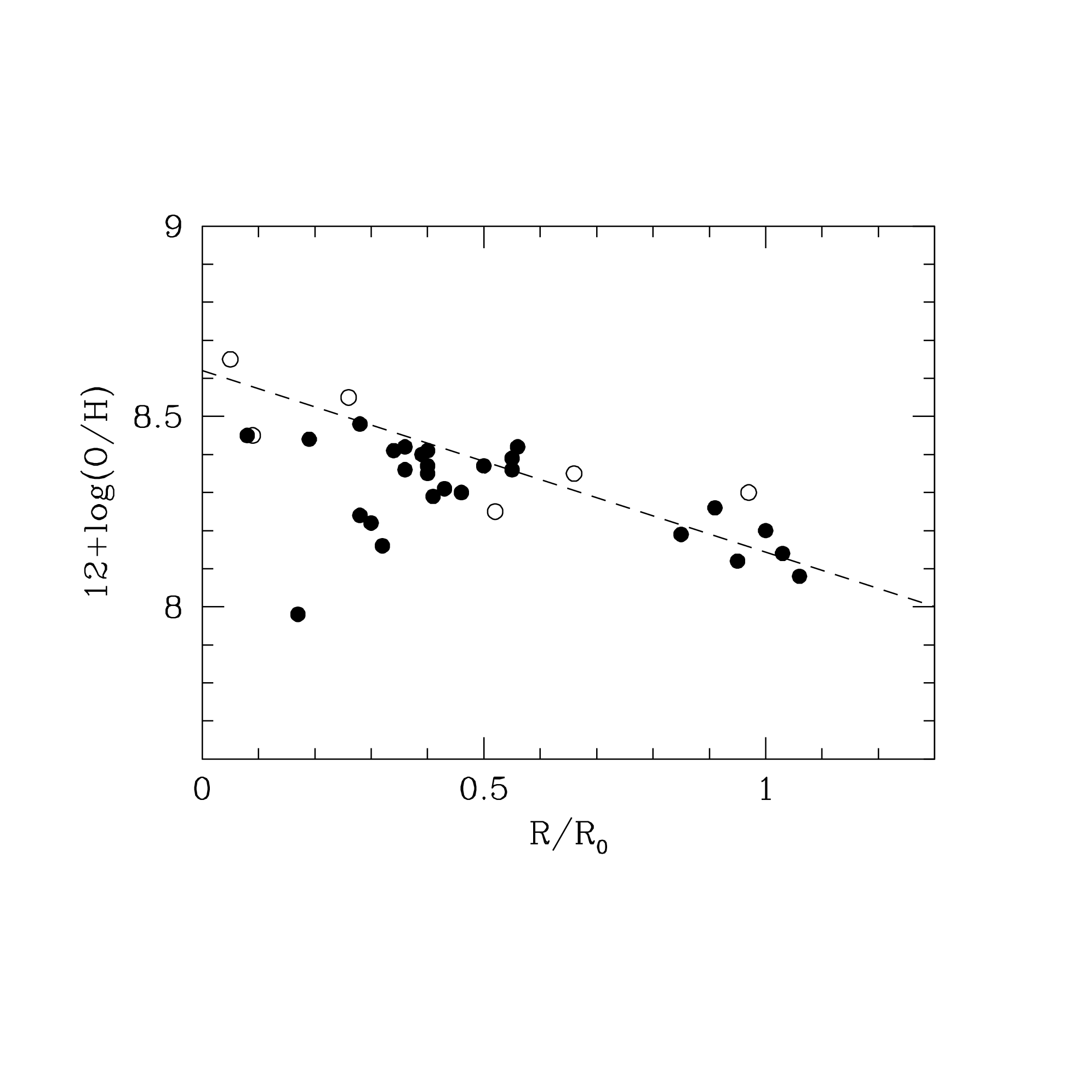} 
% \vspace*{-1.0 cm}
 \caption{The radial oxygen abundance gradient in NGC~300 measured from the six B supergiants analyzed by Urbaneja et al.~(2005a: open circles), the A-type supergiants studied by Kudritzki et al.~(2008: only the fit to their data is shown here by the dashed line), and the \hii\/ regions observed by Bresolin et al.~(2008, in prep., and this work: full circles). The radial coordinate is in units of the isophotal radius. The nebular abundances have all been determined from the strength of the \oiii\lin4363 auroral line.}
   \label{fig1}
\end{center}
\end{figure}

A variety of massive star indicators have been studied in NGC~300, including the blue supergiants (\cite[Bresolin et al.~2002]{}; \cite[Kudritzki et al.~2008]{}), Cepheids (\cite[Gieren et al.~2005]{}), W-R stars (\cite[Schild et al.~2003]{}; \cite[Crowther et al.~2007]{}) and SN remnants (\cite[Pannuti et al.~2000]{}; \cite[Payne et al.~2004]{}). The oxygen abundances of 6 B-type supergiants have been determined by \cite[Urbaneja et al.~(2005a)]{}. The only nebular line emission information available until recently was based on spectra that were obtained with 4m-class telescopes (e.g.~\cite[Deharveng et al.~1988]{}), and in which the temperature-sensitive auroral lines remained undetected. The comparison between stellar and nebular abundances has therefore been complicated by the assumptions made about the calibration of the strong-line methods used to derive oxygen abundances for the \hii\/ regions. The situation has been recently improved by the author with the acquisition  of deep optical spectra of \hii\/ regions in NGC~300 with the ESO Very Large Telescope. The faint \oiii\lin4363 line has been measured in some 20 regions. The preliminary results obtained for the oxygen abundance distribution across the disk of NGC~300 agrees very well with the stellar abundances of \cite[Urbaneja et al.~(2005a)]{} and with the A-type supergiant metallicities measured by \cite[Kudritzki et al.~(2008; see also Kudritzki's contribution for this conference)]{}, as shown in Fig.\,\ref{fig1}.

\subsection{M33}

Considerably more observational work is available for M33 than for NGC~300, especially concerning the ionized gas. Recent work has been carried out on the \hii\/ regions by \cite[Crockett et al.~(2006)]{} and \cite[Magrini et al.~(2007b)]{}.
The slope of the nebular oxygen abundance gradient agrees with the B supergiant value by \cite[Urbaneja et al.~(2005b)]{}, but in this case a $\sim$0.2 dex offset exists (stellar value is higher). It is also worth noticing that (a) the beat Cepheids studied by \cite[Beaulieu et al.~(2006)]{} provide oxygen abundances that are somewhat intermediate between those found for the \hii\/ regions and the B supergiants; (b) the slope from the recent publication by \cite[Rosolowsky \& Simon~(2007)]{} is about half the value from the works mentioned above; (c) 
the oxygen abundance determined by \cite[Esteban et al.~(2002)]{} from metal recombination lines for the giant \hii\/ region NGC~604 is about 0.2 dex higher than the value determined from collisionally excited lines, and is more in agreement with the B supergiant abundances. In this regard, it is important to note that in the Orion nebula, the three B dwarfs analyzed by \cite[Simon-Diaz et al.~(2006)]{} provide an oxygen abundance that agrees with the nebular value based on metal recombination lines by \cite[Esteban et al.~(2004)]{}, if the depletion by dust, estimated at 0.08 dex by Esteban et al., is neglected.

\section{Hot spots}
In the nearby universe
about one spiral galaxy out of five possesses circumnuclear rings of star-formation (\cite[Knapen~2005]{}). These sites of massive star formation, known as `hot spots', are important to understand the processes that lead to inflows of gas into the central regions of barred spiral galaxies, and the connection between bars and circumnuclear rings.
Hot spots are ideal places to study massive star formation in high-metallicity environments, and to understand the effects of bars, merging events and interactions between galaxies on the star formation process.

In their study of NGC~7742 \cite[Mazzuca et al.~(2006)]{} have provided a remarkable example of the power of integral field spectroscopy in the detailed investigation of the physical conditions in the nuclear regions of spiral galaxies. Their maps of emission line ratios have shown that the main source of ionization in the ring of hot spots is photoionization, while shocks dominate just outside the ring. The metallicity deduced from the \nii/\sii\/ diagnostic is approximately solar. A similar conclusion on the chemical abundance has been drawn by \cite[Sarzi et al.~(2007)]{} for a sample of eight spirals with circumnuclear star forming activity. Using instead a strong-line method, calibrated with a sample of extragalactic \hii\/ regions with auroral line detections, \cite[Diaz et al.~(2007)]{} have measured 
central abundances for three hot spot galaxies in the range 12+log(O/H)=8.60--8.85 , i.e. between solar and 1.5$\times$ solar. These values are therefore not very high, and do not reach the 2-3$\times$ solar values that were quoted just a few years ago. N/O abundance ratios are found to lie between 2 and 4 times the solar value.

The study of the gas kinematics in the circumnuclear regions of spirals is also of particular interest. Evidence for non-circular motions of the ionized gas has been found in NGC~3351 by \cite[Hagele et al.~(2007)]{}, consistent with an infall of gas at 25 km/s. A similar result has been obtained by \cite[Allard et al.~(2006)]{} in M101. These authors have also found evidence for the presence of azimuthal age gradients along the circumnuclear ring of this galaxy. The youngest massive stars are located at the contact points (resonances) between the ring and the dust lanes, where the gas inflow from the disk, under the action of the bar, intersects the circumnuclear ring.

The star formation history of the circumnuclear rings in eight different hot spot galaxies has been studied by \cite[Sarzi et al.~(2007)]{}, based on the comparison of absorption line indices with population synthesis models. Their conclusion is that the star formation in the rings occurs in episodes extending over hundreds of Myr, thus excluding continuous star formation as well as single burst events. The circumnuclear rings are therefore rather stable features.
The models by Sarzi et al.~also allow for the build-up of the metallicity to approximately the solar value, as well as of the stellar mass in the central regions of spiral galaxies. 

No hot spots with spectroscopic signatures from W-R stars have been detected by Sarzi et al.~(2007) in their sample, perhaps as a consequence of the fact that the bulk of the massive stars are formed over extended periods, rather than in short bursts. However, at this conference Hattori and collaborators have presented evidence for a large number of W-R stars in the circumnuclear hot spots in the starburst galaxy NGC~7469. W-R spectroscopic features have been found by \cite[Bresolin \& Kennicutt~(2002)]{} in a nuclear hot spot of M83. The metallicity measured from auroral lines by \cite[Bresolin et al.~(2005)]{} is approximately 1.6$\times$ solar, or 12+log(O/H)=8.94. The central region of M83 is interesting from other points of view. From the analysis of the velocity field \cite[Diaz et al.~(2006)]{} have found evidence for the presence of a hidden mass concentration ($\sim$1.6$\times$10$^7$\,\msun) near, but not coincident with, the optical nucleus of the galaxy. This mass lies at the end of a nuclear star-forming arc, whose size is compatible with the dynamical crossing time of the nuclear region. Along this arc the ages of the young clusters decrease towards the mass concentration. Diaz at el. propose that this hidden mass represent either bar-driven material funneled into the central regions of the galaxy, or the remnant of a recent merger event. We could therefore be witnessing an ongoing process that can eventually lead to the formation of a collapsed object and perhaps to the onset of AGN activity in a spiral galaxy at our galactic doorstep.

\begin{discussion}

\discuss{Dopita}{Given the disagreement between the various ways of determining the metallicity from H {\scshape ii} regions and their comparison with stars, it might be timely to return to the method I used with D'Odorico and Benvenuti many years ago - to use evolved SNR. The new shock models are now much better, and the observation of individual SNR in nearby spiral galaxies is much easier than 20 years ago.}

\discuss{Bresolin}{I agree, and once again NGC~300, where many SNR are known from recent observational work, can offer us a good starting point for this kind of investigation.}

\discuss{Herrero}{Fabio, I have a comment and a question. Comment: Esteban et al. do not derive the O content of the dust in Orion, but derive it by comparing the results by Cunha and Lambert (1994) for stars in Orion with their own results for the gas phase. Our results for B stars in Orion agree with those of Esteban et al. for the gas phase. There is no need to assume any O depletion into dust. Question: You have shown some preliminary results from your recent work about abundances in H {\scshape ii} regions in NGC~300. One of the H {\scshape ii} regions is very close to the center of the galaxy and yet has an oxygen abundance that is extremely low. How accurate is that value? Is there something peculiar about that H {\scshape ii} region or is it spectrum remarkable?}

\discuss{Bresolin}{First, I need to remind you that the results I presented for NGC~300 are still very preliminary. In the coming months I will work to produce more definitive values for the chemical abundances in the H {\scshape ii} regions I studied. I can expect some variations in some of the individual values  once the proper analysis is completed, although I do not think that the general trend and the good agreement with the stars will change. Second, I think that it is perhaps a little naive to expect that the chemical composition of a single H {\scshape ii} region or star always compares very well with that expected from its galactocentric position and the radial abundance gradient. The intrinsic scatter in abundance can also be rather large, and some peculiarites of a few individual H {\scshape ii} regions and/or stars can also be expected.} 

\end{discussion}

\end{document}